\documentclass[aps,prl,amsmath,twocolumn,showpacs,amssymb,superscriptaddress]{revtex4}
\usepackage{graphicx} 
\usepackage{color}
\usepackage{times}

\usepackage{bm}

\begin{document}

\title{Basin of Attraction Determines Hysteresis in Explosive Synchronization}

\author{Yong Zou}
    \affiliation{Department of Physics, East China Normal University, Shanghai,
200062, China}
    \affiliation{Potsdam Institute for Climate Impact Research, P.O. Box 601203,
    14412 Potsdam, Germany}
\author{Tiago Pereira}
    \affiliation{Department of Mathematics, Imperial College London, London SW72AZ, United Kingdom}
    \affiliation{London Mathematical Laboratory, London WC2N 6DF, United
    Kingdom}
\author{Michael Small}
	\affiliation{School of Mathematics and Statistics, University of Western
Australia, Crawley, WA 6009, Australia}
\author{Zonghua Liu}
	\affiliation{Department of Physics, East China Normal University, Shanghai,
200062, China}
\author{J\"urgen Kurths}
	 \affiliation{Potsdam Institute for Climate Impact Research, P.O. Box 601203,
 14412 Potsdam, Germany}
	
\date{\today}

\begin{abstract}
Spontaneous explosive emergent  behavior takes place in heterogeneous networks
when the frequencies of the nodes are positively correlated to the node degree.
A central feature of such explosive transitions is a hysteretic behavior at the
transition to synchronization. We unravel the underlying mechanisms and show
that the dynamical origin of the hysteresis is a change of basin of attraction
of the synchronization state. Our findings hold for heterogeneous networks with
star graph motifs such as scale free networks, and hence reveal how microscopic
network parameters such as node degree and frequency affect the global network
properties and can be used for network design and control.
\end{abstract}

\pacs{05.45.Xt, 89.75.Hc, 05.45.Ac}
\maketitle

Emerging abrupt transitions are ubiquitous in complex systems, and play a
crucial role in human society and a wide variety of fields
\cite{Dorogovtsev_RMP2008}. In particular, abrupt transitions to synchronization
in networks with heterogenous degree distribution have attracted much attention.
Previous works suggest that such transitions are due to a positive
correlation between the frequency and degree of the node
\cite{JesusPRL2011,Peron_PhysRevE2012,Skardal_EPL2013,Wuye_EPL2013,Coutinho_PRE2013,Li_PRE2013,Leyva_SciRep2013,Zhang_PRE2013}.
Abrupt transition has been observed in scale-free (SF) networks
\cite{JesusPRL2011}, electronic circuits \cite{LeyvaPRL2012}, time delayed
systems \cite{PeronPRE2012}, and a second order Kuramoto
model~\cite{Ji_PRL2013}.

A central feature of these emerging abrupt transitions is a hysteretic behavior
at the onset of synchronization. As the interaction strength is increased
adiabatically, the network experiences a fast explosive jump from an incoherent
state to a coherent one. Moreover, there is a sudden drop from the coherent
state to the incoherent one when the coupling strength is progressively
decreased in the backward direction. These two curves (called forward and
backward continuation below, respectively) do not overlap, instead, showing a
hysteretic behavior. The hystereris in abrupt transitions is due to the network
interaction and hence opens new paradigms for network control as coherence and
incoherence coexist in the hysteresis loop. Despite this great interest,
hysteresis at the transition to synchronization remains elusive. In
particular, it is unclear on a microscopic level how network parameters affect
the critical coupling thresholds (the hysteresis loop) and what the dynamical
origins are for the hysteresis associated with explosive synchronization.

In this Letter, we investigate hysteresis associated with the explosive
transition scenario first in networks with a star graph motif,  and then in generic SF
networks. Our results reveal that correlation in frequency-degree leads to the
existence of a phase locking state and that the hysteretic behavior is
attributed to the basin of attraction of phase locking. The phase locking state
ceases to exist at a critical parameter $\lambda_c^b$, corresponding to
synchronization loss coming from a coherent state. Starting from an incoherent
state and moving toward coherence, our analysis suggests that the locking
manifold changes its basin of attraction at a critical parameter $\lambda_c^f$
and the locking manifold becomes globally attractive. We find that whereas the
backward coupling threshold $\lambda_c^b$ tends to a constant value for large
networks, the forward critical $\lambda_c^f$ scales with the system size.

In a heterogeneous network such as SF networks,  hubs play a dominant role for
both structural organization \cite{JuddSmallStemler} and dynamical processes \cite{Judd2013}, e.g., providing
substantial resilience for preventing cascading failures
\cite{gallos_PRL2005,Motter_PRL2003,Pereira_PRE2010}. Hubs are modeled as star
motifs. A star is composed of $K$ ($K \geq 2$) peripheral nodes (or leaves)
connected to the hub. Let us start by keeping the same setting for the
frequency-degree correlation as initially explained in~\cite{JesusPRL2011}. The
hub has a frequency $\omega_{K+1} = K \omega$, while all the leaves
have the same frequency $\omega_j = \omega$ for $1\le j \le K$. Later on, we
will generalize to non-identical leaf nodes. The equations of motion are
{\small
\begin{align}\label{eq:hub}
 \dot{\phi}_{K+1} & = K \omega + \lambda \sum_{j=1}^{K} \sin(\phi_j -
 \phi_{K+1}), \\ \label{eq:leaf} \dot{\phi}_{j} & = \omega + \lambda
 \sin(\phi_{K+1} - \phi_j), \;\mbox{ for }  1\le j \le K,
\end{align}}
\noindent 
where $\phi_{K+1, j}$ are phase dynamics of the hub and leaf nodes,
respectively, $\lambda$ is the coupling strength. The Kuramoto order parameter
$\mathcal{R}(t)$ is defined as {\small $ \mathcal{R}(t) e^{i \Psi(t)} = 
\sum_{j=1}^{K+1} e^{i\phi_j}  / (K+1)$}. We quantify coherence by $r = \left <
\mathcal{R}(t) \right >_{T}$, where $\left< \cdot \right >_{T}$ denotes a time
average with $T \gg 1$. Small values of the parameter $r$ indicate incoherent
behavior. In contrast, as $r \rightarrow 1$ we encounter a highly coherent
state.

{\it Main Results:} The backward critical coupling $\lambda_c^b$ and the forward
critical coupling $\lambda_c^f$ are determined by respectively local and global
attractivity properties of a locking manifold $M_a$. Hence, the basin of
attraction of $M_a$ governs onset of hysteresis. Moreover, the scaling
relationships of the coupling thresholds on the degree $K$ are respectively
given by $\lambda_c^b \rightarrow \omega$, and $\lambda_c^f \propto K^{1/2}
\omega$, for $K \gg 1$.  Our results are based on the theory of invariant
manifolds, and the recent new findings about persistence of synchronization
\cite{Pereira_PRL2013,Pereira_xiv2013}, together with
the attractivity and basin of attraction \cite{Menck_NatPhys2013}. 

{\it Backward Continuation -- From Coherence to Incoherence: } 
We start from a coherent state where the nodes are phase locked and decrease the
coupling until we obtain a loss of coherence. We perform a local stability
analysis to explain this scenario. The state space of
Eqs.~(\ref{eq:hub},\ref{eq:leaf}) is the $K+1$ dimensional torus
$\mathbb{T}^{K+1}$. Consider $\bm{\Phi} = (\phi_1, \dots , \phi_{K+1})$, and
$\bm{\Omega}_k = (\omega, \dots,\omega, K\omega)$. Moreover, consider $\bm{H}:
\mathbb{T}^{K+1} \rightarrow \mathbb{T}^{K+1}$ defined by $\bm{H}(\bm{\Phi}) =
(\sin(\phi_{K+1} - \phi_1),\sin(\phi_{K+1} - \phi_2),\dots, \sin(\phi_{K+1} -
\phi_K),\sum_{j=1}^K \sin(\phi_{j} - \phi_{K+1}))$. With this notation the
equations of motion Eqs.~(\ref{eq:hub},\ref{eq:leaf}) are rewritten in the
compact form
$
    \bm{\dot{\Phi}} = \bm{\Omega}_k + \lambda \bm{H}(\bm{\Phi}).
$
The locking manifold is defined by 
$$ M_a := \{ \bm{\Phi} \in \mathbb{T}^{K+1} :
\phi_1 = \cdots = \phi_K \mbox{ and } \phi_{K+1} - \phi_1 = a \}. $$ 
Notice that the non-zero value of $a$ determines the locking between the hub and
the leaves. We show the existence conditions for $M_a$.

Solution curves in  $M_a$ read as 
$ 
\bm{\dot{\Phi}} = \bm{\Omega}_k - \lambda \bm{H}(\bm{a}),
$
with $\bm{a} = c(1,\dots,1) + (0,\dots,0,a)$, where $c$ is a real number. The
solutions are
$
\bm{\Phi}(t) = \big[\bm{\Omega}_k - \lambda \bm{H}(\bm{a})\big]t + \bm{\Phi_0},
$
where $\Phi_0 \in M_a$, and satisfy the condition
$\phi_{K+1} - \phi_1 = a $, which yields the equation
\begin{equation} \label{eq:malocked}
	-(K-1)\omega + \lambda (K+1) \sin a = 0. 
\end{equation}
Since $\omega$, $K$ and $\lambda$ are positive, a solution exists if
$[(K-1)\omega]/[(K+1)\lambda] \le 1$, which further leads to $ 0 < a \le
\pi/2$ \footnote{The fixed point of  Eq. (\ref{eq:malocked}) in
$(0, \pi/2]$ is stable, while the other unstable fixed point in $(\pi/2, \pi)$
is neglected.}. The equality determines the critical coupling strength for the
existence of the locking manifold, which yields the critical coupling for the
backward continuation curve as
{\small 
\begin{equation} \label{lambCBack}
 \lambda_c^b = \frac{(K-1)\omega}{K+1}.
\end{equation}
}

It turns out that whenever $M_a$ exists it is locally attractive. To see this, we study
the tangent dynamics to $M_a$. Consider $\bm{\Phi} = \bm{\Psi} +
\bm{\xi}$, where $\bm{\Psi}$ is a solution curve in $M_a$.  The equation associated
with $\bm{\xi}$ reads as 
$
	\bm{\dot{\xi}} = \lambda {\cos a} L_s \bm{\xi} + \bm{R}(\bm{\xi}),
$ 
where $L_s$ is the Laplacian matrix of the star graph, and $\bm{R}$ is a
nonlinear term satisfying $\bm{R}(\bm{\xi}) \le A \| \bm{\xi} \|^2$, for some
constant $A$. The solution of the linear part can be represented as
$
	\bm{\xi}(t) = \exp \{ \lambda \cos a L_s (t  - \tau) \} \bm{\xi}(\tau).
$
Notice that $\bm{\xi} \not \in $ span $(1,\dots,1)$, otherwise it could be
absorbed in $\bm{\Psi}$. As the Laplacian is positive semi-definite with smallest
non-zero eigenvalue equal to $1$, we get
$
	\| \bm{\xi}(t) \| \le C \exp\{- \lambda \cos a (t  - \tau) \} \| \bm{\xi}(\tau) \|
$
for some constant $C>1$. This implies that whenever the manifold $M_a$ exists it is
locally stable. Moreover, since the bound is uniform $\tau$ and exponential, the
stability will persist under the nonlinearities \cite{Pereira_xiv2013}. 

If $\lambda > \lambda_c^b$ and initial conditions are given close to the locking
manifold $M_a$, the local attractivity of $M_a$ allows us to compute the
order parameter explicitly, which reads
{\small
\begin{equation} \label{r_kstar}
 r^2
     =  \frac{K^2+1}{(K+1)^2} + \frac{2K}{(K+1)^2} \sqrt{1 -
    \left(\frac{(K-1)\omega}{(K+1) \lambda} \right)^2}.
\end{equation}}
Now note that as the locking manifold ceases to exist at $\lambda =
\lambda_c^b$, the order parameter $r$ assumes a critical value
{\small
\begin{equation}
	r_c^b = \frac{\sqrt{K^2+1}}{K+1}.
\end{equation}}
The above analysis explicitly determines the behavior of $r$ in the backward
direction. The loss of coherence occurs at the point $(\lambda_c^b, r_c^b)$
\footnote{In the sense that after this point the order parameter $\mathcal{R}$ oscillates.}.

\begin{figure}
  \centering
  \includegraphics[width=\columnwidth]{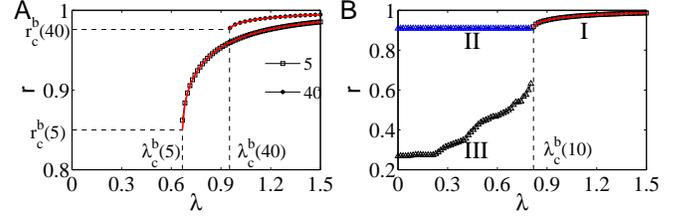}
\caption{\small {(Color online) Order parameter $r$ as a function of the
coupling $\lambda$ for various sizes. The thick lines are theoretical curves
obtained by Eq.~(\ref{r_kstar}). (A) ($K_1 = 5, K_2 = 40$), and (B) comparison
between with and without noise ($K=10$). $r$ is an arithmetic mean value of
$[\mathcal{R}_{\text{min}}(t), \mathcal{R}_{\text{max}}(t)]$ over $100$ random
realizations. Part $II$ without noise, while for parts $I, III$ there is a random
frequency mismatch $\omega_j = \omega + \zeta_j$ where $\zeta_j \in [-0.05,
0.05]$ for leaf nodes.} \label{bif_pdiff}}
\end{figure}

Figure~\ref{bif_pdiff}A shows the order parameter $r$ when the coupling is
decreased in the backward direction and starting from a coherent state. The
numerical results for various network sizes show precise agreement with the
theoretical curve given by Eq.~(\ref{r_kstar}). We obtain the critical points
$(\lambda_c^b, r_c^b)$ as predicted (denoted by coordinates in
Fig.~\ref{bif_pdiff}A ).
In the regime of $\lambda < \lambda_c^b$, the $K$ star network is reduced to two
groups: the hub and the set of leaves, evolving asynchronously.

In the next case, we consider frequency mismatches for leaves
{\small
\begin{equation} \label{eq:noiseStar}
\phi_{j}^{\prime}  = \omega + \zeta_j + \lambda \sin(\phi_{K+1} - \phi_j),
\;\mbox{ for }  1\le j \le K,
\end{equation}}
where $\zeta_j$ is a random variable uniformly distributed in $[-\varepsilon,
\varepsilon]$. Notice that if $\varepsilon$ is small and the locking manifold is
exponentially and uniformly attractive, these perturbations do not destroy the
locking manifold (part $I$ in Fig.~\ref{bif_pdiff}B). There exists another
stable locking manifold in the neighborhood of $M_a$ for $\lambda >
\lambda_c^b$. When $\lambda < \lambda_c^b$ the locking manifold no longer
exists, and as the leaves rotate at distinct frequencies, a drop in the order
parameter is observed (part $III$ in Fig.~\ref{bif_pdiff}B). In comparison, when
no noise is introduced in Eq.~(\ref{eq:noiseStar}), we find the absence of the
sudden drop in $r$ which takes place as all leaves are identical (shown by part
$II$ in Fig.~\ref{bif_pdiff}B).

\begin{figure}
	\centering
	\includegraphics[width=\columnwidth]{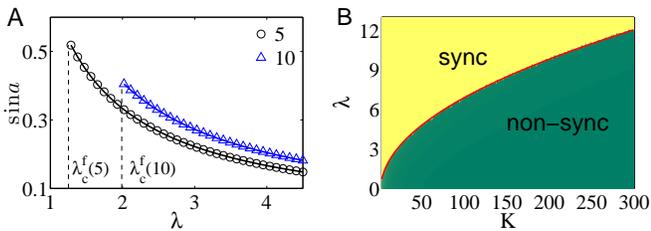}
\caption{\small {(Color online) (A) Phase difference $\sin a = \sin (\phi_{K+1}
- \phi_1)$ between the hub and the first leaf as a function of $\lambda$ for $\lambda>\lambda_c^f$.
Both circles and triangles represent the numerical simulation and the thick
lines are theoretical prediction provided by Eq. (\ref{eq:malocked}). Network size: $K=5
(\circ)$, $K=10 (\vartriangle)$. (B) Order parameter $r$ on the parameter space of
$(K, \lambda)$ for the forward continuation. The (red) thick line is from fitting the
scaling relation provided by Eq. (\ref{lambda_cf}), where the parameter $1/B =
0.6989$. } \label{aGlobalvsLambda}}
\end{figure}

{\it Forward Continuation --  From Incoherence to Coherence: } 
Starting from an incoherent state ($r$ close to zero) and increasing the
coupling strength leads to a transition towards coherence at a coupling
threshold $\lambda_c^f > \lambda_c^b$. Our numerical investigations reveal that
this behavior is related to the basin of attraction of the locking manifold
$M_a$. 
At the first stage the locking manifold $M_a$ is only locally attractive,
i.e., for $\lambda \in (\lambda_c^b,\lambda_c^f)$. Then for $\lambda>
\lambda_c^f$ the locking manifold is globally attractive, which means that
starting from an incoherent state for $\lambda > \lambda_c^f$ the network
dynamics are attracted to $M_a$. In other words, for $\lambda > \lambda_c^f$ the
phase difference is $ \phi_{K+1} - \phi_j = a$, where $a = a(\lambda)$ given by
Eq.~(\ref{eq:malocked}). We reveal that this is indeed the case, as shown in
Fig.~\ref{aGlobalvsLambda}A for distinct network sizes.

\begin{figure}
	\centering
\includegraphics[width=\columnwidth]{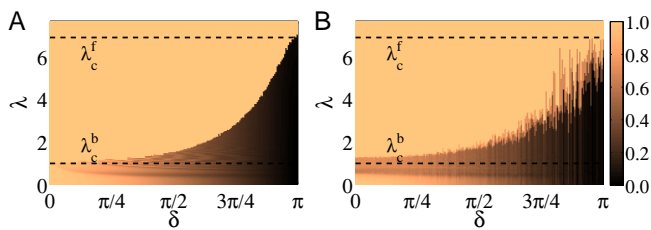}
\caption{\small{(Color online) Order parameter $r$ on the parameter space
$(\delta,\lambda)$ for the forward continuation ($K = 100$). Initial conditions
are randomly drawn from an interval $[-\delta,\delta]$. The horizontal dashed
lines are critical couplings $\lambda_c^b$ and $\lambda_c^f$ from the theory.
(A) without noise effect, (B) with frequency mismatches for leaf nodes, namely,
$\omega_j = \omega + \zeta_j$ where the random value $\zeta_j \in [-0.01,
0.01]$. \label{ColorMap}}}
\end{figure}

To analyze the basin of attraction we draw initial conditions randomly from an
interval $[-\delta,\delta]$ with $\delta \le \pi$.  Hence, if $\delta$ is close
to $\pi$ the oscillators start from an incoherent state, in contrast, if
$\delta$ is close to zero all oscillators start at a coherent state. Hence, the
value of $\delta$ enables us to capture the basin of attraction of $M_a$.  For
each pair $(\delta, \lambda)$ we compute the order parameter $r$ and the result
is shown in Fig.~\ref{ColorMap}. Note that for $\lambda < \lambda_c^b$ and small
values of $\delta$ (e.g., $\delta = 0$), the order parameter is close
to one (shown by the bright area in Fig.~\ref{ColorMap}), as all leaves are
synchronized forming a group against the hub. In this regime, $r \approx (\sin
\delta) / \delta$ \footnote{The initial conditions $\phi_j^{0}$s are drawn
independently from a uniform distribution $g(\phi)$ supported in
$[-\delta,\delta]$. If $K$ is large the law of large number yields $r =
\frac{1}{2\delta} \int_{-\delta}^{\delta} e ^{i \phi} d\phi = \sin \delta /
\delta.$},  explaining why high values of $r$ is observed for small $\delta$.

For $\lambda > \lambda_c^f$ and $\delta = \pi$, the oscillators start at an
incoherent state and then tend to the locking manifold $M_a$ leading $r$ to be
close to $1$. This scenario is not affected by the presence of mismatches in the
oscillator frequencies as shown by Fig.~\ref{ColorMap}B. 

While the value $\lambda_c^b$ tends to a constant, the critical forward coupling
$\lambda_c^f$ scales roughly as $K^{1/2}$. Hence, for large $K$ the difference
between the forward and backward coupling thresholds becomes severe. Note that
our results below correspond to $\delta = \pi$, namely, the initial conditions
are randomly chosen from $[-\pi, \pi]$. We calculate the order parameter $r$ for
various network sizes, $K \in [3, 300]$ for each coupling strength $\lambda$,
yielding a color coded parameter space of $(K, \lambda)$ as shown in
Fig.~\ref{aGlobalvsLambda}B, where we observe an abrupt transition from an
incoherent to a coherent state.

To obtain an analytical understanding of this scaling property for
$\lambda_c^f$, we use the theory recently developed in Ref.
\cite{Pereira_PRL2013}. To this end we write this phase locking problem between
the hub and leaves as a perturbation of an identical synchronization problem.
Hence, the isolated dynamics of the hub reads as $ \dot{\phi}_{K+1}= \omega +
g_{K+1}$. Representing Eqs.~(\ref{eq:hub},\ref{eq:leaf}) in block form yields the perturbation $\bm{G} =
(0,\dots,0,(K-1) \omega)$. The block equation then reads
$
    \bm{\dot{\Phi}} = \bm{\Omega} 
     + \bm{H}(\bm{\Phi}) + \bm{G},
$
where $ \bm{\Omega} = (\omega, \cdots,\omega)$. After an involved algebraic
manipulation following Refs. \cite{Pereira_PRL2013,Pereira_xiv2013}, we obtain
$
 (\sum_{j} |\phi_{K+1} 
  - \phi_j |^2 )^{1/2}  \le  (\sigma \| \bm{G} \| )
 /  \lambda,
$
where $\| \cdot \|$ denotes the Euclidean norm, and $\sigma$ is a constant.

Hence, starting from an incoherent state to obtain a coherent one if the
trajectories enter the neighborhood of a fully synchronized state $\phi_1 =
\cdots = \phi_{K+1}$.  This neighborhood contains the locking manifold $M_a$ as
$a$ tends to zero. Using the above bounds for the phase difference we obtain the
scaling behavior of the coupling parameter. Indeed, notice that $ \| \bm{G} \| =
(K-1) \omega$ and using that the oscillators start from a incoherent state $
|\phi_{K+1} - \phi_j | < 2 \pi,$ we obtain $
( \sum_{j} |\phi_{K+1} - \phi_j |^2 )^{1/2}  \le 2 \pi \sqrt{K}.
 $
Manipulating this equation, we obtain that the coupling strength scales $\lambda
\propto [(K-1)\omega] / \sqrt{K}$. This coupling corresponds to the necessary to
get coherence starting from an incoherent state. However, this is precisely the
forward coupling strength $\lambda_c^f$. Hence, trajectories will approach the
locking manifold with the coupling strength
\begin{equation}\label{lambda_cf}
	\lambda_c^f \approx	\Big(\frac{K-1}{\sqrt{K}} \frac{1}{B} \Big)\omega
\end{equation}
\noindent
where $B$ is a constant parameter \footnote{This constant may depend on how we
choose the initial conditions. There are two ways to choose initial conditions
(ICs). (i) The ICs for the coupling $\lambda + \Delta\lambda$ are the final
states when coupling equals to $\lambda$ as suggested in \cite{JesusPRL2011},
which yields $1/B = 0.6989$. (ii) The ICs are independent randomly chosen from
the interval $[-\pi, \pi]$ for each coupling $\lambda$, resulting in $1 / B = 2
/ \pi$.}. Our numerical result in Fig. \ref{aGlobalvsLambda}B shows an excellent
agreement with this theoretical analysis.

{\it Scale-Free Networks:} 
The results above for star graphs can be straightforwardly applied to explain
the recent findings of hysteresis in SF networks when the mean degree is small,
due to the role of hubs. If the average degree of the network is small, then the
network can be seen as a collection of star graphs. In particular this can be
seen to be the case for {\em random} power-law graphs as the exponent
$\gamma\rightarrow 3$ \cite{JuddSmallStemler}. More generally, the role of hubs
in scale free networks is certainly dominant, and it is only for low values of
$\gamma$ (i.e. $\gamma <2.5$) that one will expect graphs with more links than a
tree and hence exhibiting an excess of loops and significant deviation from a
composition of hubs \cite{JuddSmallStemler}. Of course, experimentally observed
SF networks are very rarely trees --- nonetheless, they remain defined by their
high degree hubs.

In such networks, each hub and its corresponding neighboring nodes of low
degrees will have a locking manifold, and the connections between low degree
nodes of distinct hubs act as small perturbations. Therefore, the investigation
of the hysteresis-like behavior on a SF network can be greatly explained by a
star graph with frequency mismatches for leaf nodes (i.e., Eq.
(\ref{eq:noiseStar})). In combination with the global order parameter $r$, it is
convenient to compute the local order parameter $r_i$ for the $i$-th hub.
Parameter $r_i$ is obtained by averaging only over nodes connected to the $i$-th
hub.  The local order parameters play a role in the SF network as hubs are
connected to a different number of nodes $K_i$.

\begin{figure}
  \centering
  \includegraphics[width=\columnwidth]{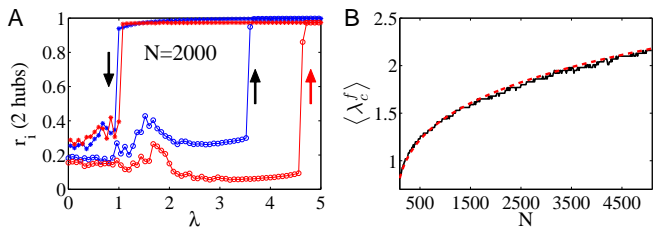}
\caption{\small {(Color online) (A) Local order parameter $r_i$ versus coupling
strength for two chosen hubs ($K_1=39 \; \text{red}, K_2= 24 \; \text{blue}$;
$\lambda_{c,1}^f = 4.64$, $\lambda_{c,2}^f = 3.63$). (B) Critical coupling
$\left < \lambda_c^f \right >$ vs. sizes of SF networks (dashed line is the
theoretical curve predicted by Eq. (\ref{lambda_cfBAnet}), where $\left<\cdot
\right>$ is an ensemble average over $50$ network realizations. }
\label{localOrderHubs}}
\end{figure}

We generate a SF network by means of the Barab\'asi-Albert model with $m_0 = 1$
\cite{Albert2002}\footnote{We observe that the unbiased network generation model
of \cite{JuddSmallStemler} actually generates {\em more} hub-like networks than
the Barab\'asi-Albert algorithm and hence, we are considering a relatively
difficult test case.}. We analyze the hubs by the local order parameter $r_i$.
As predicted by Eq. (\ref{lambCBack}), the backward continuation for various
hubs of different sizes converges to the critical value $\lambda_c^b \to \omega,
K \gg 1$ (size independent shown by the backward curves of the two largest hubs
in Fig. \ref{localOrderHubs}A). In contrast, since hubs often do not have the
same degrees, the local order parameter $r_i$ will present forward transitions
at distinct coupling values, but still governed by Eq. (\ref{lambda_cf}). In
Fig. \ref{localOrderHubs}A, we show the forward curves for the two largest hubs
of  a network with 2000 nodes with degrees $K_1 = 39$ and $K_2 = 24$. Denote
$\lambda_{c,1}^f$ the critical value of the largest hub, and $\lambda_{c,2}^f$
for the second largest. Our results predict that $\lambda_{c,1}^f$/
$\lambda_{c,2}^f = \sqrt{K_1/K_2} = 1.275$, which is in an excellent agreement
with our simulations yielding $\lambda_{c,1}^f$/ $\lambda_{c,2}^f  = 1.278$.
This result on the dominant role of hubs holds for networks of various sizes and
random realizations.

We calculate the forward critical coupling $\lambda_c^f$ for various network
sizes. For one network of size $N$, we numerically estimate the threshold
$\lambda_c^f$ by fixing a level of coherence over hubs (say $r = 0.5$ over the
top 20 hubs). In addition we consider the expectation of $\langle \lambda_c^f
\rangle$ with respect to the network ensemble. The dependence of the expected
coupling $\langle \lambda_c^f \rangle$ on the system size follows: Note that the
expected degree of the largest hub $K_{max}$ scales as $N^{1/(\gamma - 1)}$
\cite{Mori2005}, which means that on average the hubs are star motifs with
$N^{1/(\gamma - 1)}$ leaves. Our previous considerations show that
\begin{equation} \label{lambda_cfBAnet}
\langle \lambda_c^f \rangle \propto N^{\frac{1}{2 (\gamma - 1)}}.
\end{equation}
This is in agreement with our numerical experiments on the SF
network where $\gamma = 3$, as shown in Fig. \ref{localOrderHubs}B. 

Nonetheless, if the mean degree is high enough (for large $m_0$), the leaves of
the hubs will strongly interact. So the modeling of a SF network as a collection
of stars is no longer useful. In such situations mean field approaches may
capture the behavior of the leaves \cite{JesusPRL2011,PeronPRE2012}. An
interesting question is when the crossover between our approach and the mean
field scheme takes place. Judd \cite{Judd2013} provides strong indication that
even when the approximation is not precise,  modelling of a SF network as a
collection of stars may still be useful --- and indicates that SF networks which
are collections of stars are actually quite common\cite{JuddSmallStemler}.

In summary, we have shown that the abrupt transition in the Kuramoto model in
both star motifs and SF networks is associated with a locking manifold and its
local and global attractivity properties. The critical coupling associated with
loss of coherence is determined by the existence of a locking manifold, whereas
the critical coupling responsible for attaining coherence starting from an
incoherent state is related to a change in the basin of attraction of the
locking manifold. We have uncovered the distinct dependence of both coupling
thresholds on the network size, revealing that the hysteresis is stronger in
large networks. Our findings provide methods to control the transition and
hysteresis in terms of microscopic network parameters.

This work was partially supported by the NNSFC (11305062, 11135001), SRFDP
(20130076120003) (YZ, ZL), Australian Research Council Future
Fellowship FT110100896 (MS), Marie Curie IIF Fellowship (303180), CNPq, and Leverhulme
Trust Grant (RPG-279) (TP).


\end{document}